\documentclass{ws-ijmpa}
\usepackage[super,compress]{cite}
\usepackage{graphicx}
\begin{document}
\markboth{Dimitri Bourilkov}{Machine and Deep Learning Applications in Particle Physics}

%
\catchline{}{}{}{}{}
%

\title{Machine and Deep Learning Applications in Particle Physics}

\author{Dimitri Bourilkov}

\address{Physics Department, University of Florida, PO Box 118440\\
Gainesville, Florida 32611, USA\\
dimi@ufl.edu}

\maketitle

\begin{history}
\received{25 October 2019}
\end{history}

\begin{abstract}
The many ways in which machine and deep learning are transforming the
analysis and simulation of data in particle physics are reviewed. The
main methods based on boosted decision trees and various types of
neural networks are introduced, and cutting-edge applications in the
experimental and theoretical/phenomenological domains are
highlighted. After describing the challenges in the application of
these novel analysis techniques, the review concludes by discussing
the interactions between physics and machine learning as a two-way
street enriching both disciplines and helping to meet the present and
future challenges of data-intensive science at the energy and
intensity frontiers.
\end{abstract}

\ccode{PACS numbers: 07.05.Mh}


\section{Introduction}

The twenty-first century has brought widespread advances in the
natural and social sciences by making them data-intensive. The
rise in computing power and networking has allowed to amass ever
expanding collections of data in the petabyte and even exabyte
range~\footnote{
For pioneering developments in 2001-5 see e.g. the International
Virtual-Data Grid Laboratory for Data Intensive Science (iVDGL),
combining the efforts of the Laser Interferometer Gravitationalwave
Observatory (LIGO), the ATLAS and CMS detectors at LHC at CERN and the
Sloan Digital Sky Survey (SDSS)~\cite{iVDGL}.}.  The progress in
social media and e-commerce has only added to the flood. This in turn
has accelerated the development of novel techniques needed to analyze
the data and extract useful and timely information from it. The field
of data science was born.

The traditional way to analyze, or generate simulated, data is to
first develop algorithms based on domain knowledge, then implement
them in software, and use the resulting programs to analyze or
generate data. This process is labor intensive, and analyzing complex
datasets with many input variables becomes increasingly difficult and
sometimes intractable. Artificial intelligence (AI) and the subfield
of machine learning (ML) attack these problems in a different way:
instead of humans developing highly specialized algorithms, computers
learn from data how to analyze complex data and produce the desired
results. There is no need to explicitly program the computers.
Instead, ML algorithms use (often large amounts of) data to build
models with relatively small human intervention. These models can then
be applied to predict the behavior of new, previously unseen data, to
detect anomalies or to generate simulated data. While early work
stretches back more than fifty years, progress was slow for long
periods of time. Advances in academic research paired with the needs
of large companies like Google, IBM, Amazon, Facebook and Netflix,
just to name a few, are producing a fundamental paradigm shift,
especially with the recent successes of deep learning (for an
excellent introduction to the topic, see e.g.~\cite{DL}).

Using mostly traditional analysis methods, physics has advanced
rapidly, establishing the Standard Model (SM) of particle physics, and
more recently its cosmological homologue, $\Lambda$CDM. The coming
years will bring unprecedented amounts of data and complexity at the
Large Hadron Collider (LHC), accelerating protons at CERN, as well as
at the intensity frontier and elsewhere. Extracting the underlying
physics in the same way becomes more and more challenging, or simply
impossible in a timely manner. That explains the recent spark of
interest in ML (for excellent recent reviews and plans for the future,
see e.g.~\cite{Radovic:2018dip,Albertsson:2018maf,Carleo:2019ptp}).

The physical sciences are in a unique position. While in many other
fields there are less firm theoretical foundations or models,
physicists have well established methods to predict and to compare the
results of experiments to theoretical calculations, as the many
successes of the SM attest. This means that physics motivated ML
methods can be developed and applied, accelerating the learning
process and making it more efficient and precise. At the same time the
breath-taking advances in data science and computing technology will
help to address the coming challenges in particle physics.

This review is not meant to be all-encompassing. Rather, some
cutting-edge applications at the energy and intensity frontiers of
particle physics are selected to illustrate the many amazing ways in
which ML is applied, and to highlight both the successes and the
challenges. The review is organized as follows: after an introduction
to ML, the applications in experimental high energy physics (HEP) are
reviewed in section 2, and in phenomenological and theoretical HEP in
section 3. Open issues and challenges are discussed in section 4,
followed by a more general overview of how ML works or can be improved
in section 5, and an outlook in section 6.

\subsection{Machine Learning Basics and Vocabulary}.

With the increasing complexity of events in high energy physics,
the importance of multivariate analysis for LHC has been recognized
before the start of data taking. The main motivation was to go beyond
the traditional methods for event selection by applying series of
cuts on individual variables, and be able to use correlations and more
intricate patterns in the multidimensional data. A
workshop~\cite{caltechmva} at Caltech in 2008 was dedicated to the
topic; ML techniques were practically not on the radar. What a sea
change ten years later.

Machine learning algorithms, which are general in nature and not
task-specific, are geared towards improving the measurable performance
on some given task by training on more and more data.
The data are split in training, validation and test subsets. The first
two are often combined together, as in cross-validation, where a
different chunk of the data is used at each training step to estimate
the predictive power of a model. The ultimate measure of the model
generalization ability is how it will perform on unseen test data,
which can include real or future data. To avoid the danger of
overfitting, in ML approximate solutions are preferred: the goal is to
learn the essential features of the data, not all the quirks and
fluctuations of the training sample; this way models will generalize
better. Instead of an exact, ``ideal'', a ``good enough'' solution is
favored, even when several runs on the same data, due to random
effects, generate several similar, but not identical models. In ML
courses often Ockham's razor, named for the fourteenth century
Franciscan friar, is cited as a helpful path to generalizibility:
``More things should not be used than are necessary.'' Based on our
knowledge about physics, we can be less restrictive. As Albert
Einstein famously said: ``Everything should be made as simple as
possible, but not simpler.'' Good ML models find a balance between the
two. Once a model is trained, it can be applied on new data, the so
called inference. Usually this step is much less computationally
intensive, providing sizable speed-ups in processing data.

Early ML applications in HEP often used decision trees: a tree like
model for decisions, starting at the root, climbing up the branches
and reaching the leaves, where each leaf represents a decision. For
classification problems, each leaf represents our decision assigning a
data item to a class (binary or multiclass problems). In HEP, the most
widely used are boosted decision trees (BDT), which convert ``weak''
to ``strong'' learners.


Artificial neural networks (ANN or just NN) try to imitate in a
simplified way biological brains. The neurons and synapses are
replaced with connected layers of nodes (units, or sometimes even
simply neurons) and edges. A node takes inputs from its connections as
real numbers (a weighted sum of the connected outputs from the
previous layer), and performs a non-linear transformation to form its
output. Typical activation functions for this are: $sigmoid$
(logistic) and $tanh$ where the output is limited below $|1|$ for any
input values, and the rectified linear unit $ReLU$ ($max(0,x)$ or the
positive part of the argument). NN have an input, an output, and one
or multiple (``deep learning''- DL) hidden layers. Deep NN are denoted
as DNN.

The learning can be supervised based on pairs of inputs with known
outputs for training, or unsupervised, for example density estimation,
clustering or compression. A cost or loss function measuring the
``distance'' between the current and the desired outcomes is minimized
to train the model. Classical optimization aims to minimize the cost
function on the available (training) data, while in ML the goal is to
generalize, or minimize the cost best, on the unseen (test) data. At
each step the weights for all the edges can be adjusted by
backpropagation based on the differentiation chain rule to reduce the
cost function by small amounts. This is the stochastic gradient
descent (SGD). The associated learning rate is similar to the
$\epsilon$ introduced by Cauchy~\cite{Cauchy} to formalize calculus in
the nineteenth century.

Many familiar terms have their equivalents in ML jargon: variables are
called features, iterations become epochs, labels often are called
targets. To speed up convergence, minimizations are carried over data
batches of limited size, and the weights adjusted, instead of
traditional global solutions in one go, which are much slower.
Multilayer architectures can be trained by backpropagation and
SGD. The fears from local minima, unwanted e.g. in HEP fit
applications, have largely dissipated. For complex phase spaces there
are many saddle points which give very similar values of the cost
function, i.e similar models~\cite{DL}. Instead of SGD, a popular
optimizer is Adam~\cite{Kingma:2014vow}, which adjusts the learning
rates per parameters and based on recent history.

While the values of edge weights are learned during training, the so
called hyperparameters, like learning rate, model architecture
(e.g. number of hidden layers and nodes per layer), activation
functions, or batch size, are set before one run of the learning cycle
begins. Depending on the data patterns to be learned or abstracted,
different values of the hyperparameters will be needed for the same ML
tool. The hyperparameter tuning necessitates several, often many
learning runs. Here is where human intervention and data scientist
skills are key.

To keep this ML overview concise, more details about specific ML
techniques will be provided throughout the text.

\section{Machine Learning in Experimental HEP}

\subsection{Classification and Event Selection}

The difficulty in extracting small signals from the towering LHC
backgrounds has helped to introduce ML techniques for classification
purposes. Classification algorithms are a type of supervised learning
where the outputs are restricted to a limited set of values, or
classes like signals or backgrounds. The Higgs analyses are a prime
example.

The discovery of the Higgs boson in 2012 by the
CMS~\cite{Chatrchyan:2012xdj} and ATLAS~\cite{Aad:2012tfa}
collaborations saw the first use of boosted decision trees in such a
high stakes search for the separation of small signals (invariant mass
peaks) over large smoothly falling backgrounds. Since then the Higgs
decays and couplings to the heavy W and Z gauge bosons, as well as the
heavy third generation quarks (bottom and top) and tau leptons, have
been observed by both ATLAS and CMS, and are consistent with the
predictions of the SM at the current level of precision. With the
Higgs boson firmly established, attention has turned to measuring its
properties.

The next frontier is observing Higgs decays and measuring its
couplings to fermions outside the third generation. The search for
Higgs decays to a pair of muons with opposite charge ($\mu^+\mu^-$)
offers the best chance to establish and measure the Higgs couplings to
the second generation. This is a very challenging undertaking: the SM
branching fraction is expected to be $\sim$0.02\%. The small
signal has to be extracted over a huge irreducible background
producing opposite sign muon pairs: Drell-Yan, top quark or W boson
pairs production. The Higgs signal has a narrow dimuon invariant mass
peak near 125~GeV, only a few GeV wide, determined by the experimental
muon momentum resolution. In contrast, the background events exhibit a
smoothly falling mass spectrum in the search region from 110 to
160~GeV.

The CMS collaboration developed a method to enhance the signal
extraction by using a BDT classifier, as implemented in the TMVA
class~\cite{Hocker:2007ht} of the ROOT analysis package~\cite{root},
augmented with automated categorization for optimal event
classification. The results of the 2016 analysis, using 35.9~fb$^{-1}$
of collision data, were published in~\cite{Sirunyan:2018hbu}. Details
of how the analysis was optimized for maximum signal sensitivity,
utilizing multivariate and machine learning techniques, are provided
in~\cite{Bourilkov1}.  Events are divided into categories based on the
transverse momentum ($p_T$) of the dimuon pair (which is higher for
the main gluon-gluon fusion signal (ggF) relative to the main
Drell-Yan background), or the presence of a high-invariant-mass dijet
pair, characteristic of vector boson fusion (VBF) signal
events. Categories are sub-divided further based on the muon
pseudorapidity ($\eta$), as central muons have better $p_T$
resolution, resulting in a sharper signal mass peak.

The training is based on one million simulated events for the various
channels, fully reconstructed in the CMS detector. Fourteen kinematic
variables characterizing the dimuon system are used, their
distributions are very similar between the signal and background
events, making the separation that much harder. The signal sample is
split into three independent sets: one for training, a second for
testing, and a third completely independent - to avoid any bias - for
the final measurement. The background samples are typically split in
75\% for training and 25\% for testing. A binary signal-background
separation is computed, yielding a BDT score between -1 and 1, where
events close to 1 are more signal-like, and events close to -1 are
more background-like.

As a last step the auto-categorizer procedure determines 15 event
categories based on $|\eta|$ and BDT scores. Performing separate
signal-plus-background fits to the data in all of these categories and
combining the results significantly increases the search sensitivity
relative to a measurement of all candidate dimuon events together. The
net result of applying machine learning techniques is a 23\% increase
in sensitivity equivalent to 50\% more data.

The ATLAS collaboration has presented an updated
result~\cite{ATLAS-Hmm} using all the Run2 data: 139 fb$^{-1}$. The
observed upper limit on the cross section times the branching ratio of
the Higgs decay to a muon pair is 1.7 times the SM prediction, so the
LHC experiments are closing in on the observation of this channel, but
will need more data.

This analysis follows a somewhat similar approach, using 14 kinematic
variables and 12 categories to optimize the separation of the signal
from the backgrounds. Data events from the sidebands and simulated
signal events enter the BDT training procedure. The XGBoost (eXtreme
Gradient Boosting)~\cite{Chen:2016btl} package is used. First a BDT is
trained in the category of events with two or more jets to disentangle
the VBF signal from the background. Three VBF categories with
different purities are defined based on this BDT score. Then the rest
of the events is divided with three BDTs providing ggF scores, and
split according to jet multiplicities with zero, one or two jets,
giving nine additional categories. To maximize the sensitivity to the
Higgs to muons decays the boundaries are adjusted by BDT scores, and
in each of the twelve BDT categories a fit to the invariant mass
spectrum from 110-–160 GeV is performed to extract the signal.

This analysis is able to achieve about 50\% higher expected
sensitivity compared to the previous ATLAS result, with roughly equal
parts due to the increase in integrated luminosity or from refinements
in the analysis techniques, where machine learning plays a key role.

A substantially more difficult task is the search for Higgs decays to
a pair of charm quarks from the second generation. The CMS
collaboration has performed a direct search for this Higgs decay where
the Higgs is produced in association with a W or Z boson, based on an
integrated luminosity of 35.9 fb$^{−1}$ collected at the CERN LHC in
2016~\cite{CMS-Hcc}.

Two types of jet topologies are analyzed: ``resolved-jet'' , where
both charm quark jets from the Higgs decay are observed, and the
``merged-jet'' topology, where the two jets from the charm quark can
only be reconstructed as a single jet. In both topologies, novel tools
based on advanced machine learning techniques are deployed.

For the ``resolved-jet'' topology BDT with gradient
boost are trained to enhance the signal-background separation. Four
categories having 0, 1 or 2 leptons from the associated W or Z decays
(the 2 lepton case subdivided depending on the p$_T$ of the vector
boson) and 25 input variables are used for training.  For the
``merged-jet'' topology a novel algorithm based on advanced ML methods
is deployed to identify jet substructures in order to tag the highly
boosted W, Z, and Higgs decays, giving sizable gains.

The use of an adversarial network~\cite{Goodfellow:2014upx} helps to
largely decorrelate the algorithm from the mass of a jet while
preserving most of its discriminating power. For example, for large
jets with p$_T\ >\ $~200 GeV, misidentification rates for charm quark
pairs of 1\%, 2.5\% and 5\% are achieved for efficiencies of 23\%,
35\% and 46\%. The corresponding b jet misidentification rates are
9\%, 17\% and 27\%.

The results of the two topologies help to provide an upper limit on
the branching ratio of Higgs decays to charm quarks. There is still a
long way to reach the sensitivity needed to observe this Higgs decay
with SM strength.

While in the early Higgs papers BDTs were the prefered ML approach,
newer analyses deploy deep learning and NN. The CMS collaboration has
measured~\cite{CMS-Htautau} the inclusive cross section for the
production and subsequent decay of Higgs bosons to tau lepton pairs
with 77.4 fb$^{−1}$ of data collected in 2016 and 2017.

A multi-classification approach is applied for each final state and
year of data-taking, eight independent tasks in total. For each of
them a fully connected feed-forward NN is trained. The architecture
consists of two hidden layers with 200 nodes each, and five or eight
nodes in the output layer, each representing an event class
prediction, depending on the final state. The total cross section as
well as cross sections for individual production modes and kinematic
regimes are obtained. This is made possible by the power of
classification using deep learning.

Another recent example is the measurement of associated production of
top quark-antiquark pairs and Higgs bosons (ttH), with Higgs decaying
to b quarks, by the CMS collaboration~\cite{CMS-ttH}. The analysis is
based on 41.5 fb$^{−1}$ collected in 2017, and combined with the 2016
analysis reaches an observed (expected) significance of 3.9 (3.5)
standard deviations above the background-only hypothesis, providing
the first evidence for ttH production with subsequent H$\rightarrow$bb
decays. Multiple classifiers are deployed, like BDTs for the dilepton
channel, or feedforward NN with three hidden layers of 100 nodes each,
as implemented in Keras~\cite{Keras}, for the single lepton channel.

The application of ML techniques for Higgs analyses at the LHC is not
a one-way street. Data from the simulations of the Higgs decays to
tau-lepton pairs were released by ATLAS to the ML community and formed
the basis for the HiggsML challenge~\cite{HiggsML}. It ran from May
to September 2014 on the Kaggle platform~\cite{HiggsMLKaggle}, was
extremely popular, attracted 785 teams with 1942 participants and
generated 35772 submissions and more than a thousand forum posts.
Probably more surprising then than now, first prize was won by Gabor
Melis, a software developer and consultant from Hungary, using
artificial NN. A special HEP meets ML award was provided to data
science graduate students Tianqi Chen and Tong He for offering the
boosted decision trees tool XGBoost, used by many participants. By now
CERN provides an open data portal~\cite{CERN-ODP} to the LHC
experiments to encourage fruitful collaboration between high energy
physicists and data scientists.


{\it Great progress in computer vision has come from convolutional
neural networks (CNN), inspired by the animal visual cortex, where
individual neurons process information only from parts of the visual
field. This ``divide-and-conquer'' strategy simplifies the NN
architecture and helps features like translational and rotational
invariance, very desirable for image recognition. Typically the
first layers of a CNN are for convolution and pooling. In a
convolution the shape of an input function is modified by another
function by taking an integral of the product of the two. The
convolutional filtering helps e.g. in edge detection.
Hand-engineering the filters is replaced by learning them from the
images. Pooling layers combine the inputs from several neurons (the
simplest being a} 2x2 {\it cluster) into one output neuron, thus
reducing the dimensionality. This is usually followed by fully
connected layers like in standard DNN for the final classification
step.}

Traditionally a HEP analysis proceeds to reconstructing higher level
objects like tracks and energy deposits in electromagnetic and hadron
calorimeters from the raw detector data, and finally arriving at
particle level objects. A promising new approach is to apply deep
learning algorithms directly to low-level detector data. This is
explored in what is called end-to-end event
classification~\cite{Andrews}. The study is based on 2012 CMS
Simulated Open Data for the decays of the Higgs boson to a pair of
photons. The gluon-gluon fusion Higgs production is the signal, while
irreducible quark fusion to photon pairs and a photon plus jet faking
a second photon events form the backgrounds in this simplified
study. The events are simulated taking into account the interactions
in the detector materials and the detailed CMS geometry.

The low level detector data is converted into images of size 170x360
in pseudorapidity $\eta$ and azimuthal angle $\varphi$ for the CMS
barrel, and two images of size 100x100 for the two CMS endcaps
extending to $|\eta|\ <\ 2.3$. Each channel contains three layers
corresponding to electromagnetic and hadron energy and track
transverse momentum. This is the electromagnetic-centric
segmentation. Alternatively, a hadron-centric segmentation with size
280x360 in $\eta - \varphi$ is used. Inspired by the recent progress
in computer vision, a CNN of the Residual Net-type
(ResNet-15)~\cite{He} is used. The initial results show promise, with
signal efficiency and background rejection on par with more
traditional approaches.

\subsection{Reconstruction}


Regression algorithms are another type of supervised learning,
providing continuous outputs which can have any numerical value within
a range. They can be deployed for reconstruction purposes in HEP
e.g. when we want to make precise determinations of continuous
quantities like hit positions, track momenta or jet energies.

At the intensity frontier advanced detectors collect record amounts of
luminosity at what would be considered ``medium'' energies by today's
standards. One example is the Beijing Electron Positron Collider
(BEPCII) running at center of mass energies 2.0--4.6 GeV. The BESIII
experiment has collected record size data samples in this
$\tau$--charm region. Advanced ML techniques have been applied for
many tasks~\cite{BESIII}. One of them is cluster reconstruction for
the cylindrical triple-GEM inner tracker, part of the 2019 upgrade to
the aging inner drift chamber. The goal is to measure the drift
cathode layer position of ionizing particles from the readouts of the
anode strips, which is the first reconstruction step. Two methods are
available: weighted by electric charge average position of the anode
strips (Q~method), or time measurement using the drift gap as kind of
micro time projection chamber (T~method). The two methods can be
combined to improve the position resolution, but this combination is
made difficult by the correlations to the incident angle. Here ML
techniques come to the rescue: a XGBoost regressor is developed to
reconstruct the initial particle positions from the Q and T
readouts. Substantial improvements over the charge centroid method are
reported.

ML techniques are entering in full force the ``sister'' field of
particle astrophysics. One development in the field of very high
energy gamma-ray astronomy is the Cherenkov Telescope Array (CTA)
which will ultimately consists of 19 telescopes in the Northern and 99
telescopes in the Southern hemisphere to cover the full sky. A
colossal amount of data in the multi-petabyte range per year is
expected. The telescope arrays are operated as a single instrument to
observe extensive air showers originating from gammas or charged
particles, and aim to separate them and measure basic characteristics
as energy, direction and impact point of the original particle. An
exploratory regression study~\cite{GammaLearn} in this direction uses
CNN with the hope to extract more information directly from the raw
data and outperform traditional approaches based on human-selected
features.

The main difficulty is that conventional CNNs are developed to process
rectangular images with regular pixel grids. The telescope outputs
here have hexagonal pixels forming hexagonal images. One, not very
satisfying approach is resampling, converting the image to a standard
one, potentially losing some information about the neighbors. This
analysis takes the more difficult route of reimplementing the
convolutional and pooling operations of CNNs by building matrices of
neighbor indices, rearranging the data accordingly and then applying
the general methods for convolution, or for pooling with different
functions depending on the task ({\it softmax, average, max}).

The next difficulty is to combine images from several telescopes to
obtain stereoscopic information. Traditional DL methods
only deal with single images, sequentially in time. This is solved by
adding a convolution block for each telescope in the array, and
feeding them all to the dense fully connected part of the network. The
exploratory study with four telescopes shows promise in the
measurements of energy, direction and impact point for incoming
particles; additional work is needed to outperform traditional methods
and solve technical details before applying the developed algorithm on
real data.

Tracking detectors form the core of most collider experiments, and
successful track reconstruction is mission critical for achieving
their goals. Reconstructing tracks is a combinatorial problem,
i.e. finding the measurements (hits) belonging to individual particles
entering the detectors from an often huge set of possible
combinations. With the transition to the High-Luminosity LHC (HL-LHC)
the complexity of this task will increase substantially. Traditional
approaches like track following (inside-out or outside-in) and Kalman
filters do not scale favorably to very high hit densities, and are
typically custom implemented for each experiment with large amount of
human efforts.

The TrackML~\cite{TrackML} project has the ambition to stimulate new
approaches and the development of new algorithms by exposing data from
a virtual, but realistic HL-LHC tracking detector to data science and
computer experts outside of the HEP community. Production of
top-antitop quark pairs is selected for the signal events, which are
then merged with 200 soft interactions (pile-up events). Fast
simulation is used to generate hits in the tracker from charged
tracks. The magnetic field is inhomogeneous, energy loss, hadronic
interactions and multiple scattering are parameterized. The silicon
tracker consists of three parts: innermost pixel detector, followed by
two layers of short and long silicon strips providing hermetic
coverage up to $|\eta|\ <\ 3$. For each collision, about ten thousand
charged particles, originating approximately from the center of the
detector, produce about ten precise hits per track in three
dimensions.

The challenge, running on the Kaggle platform~\cite{TrackMLKaggle} and
on Codalab in 2018--2019, is split in two phases: accuracy and
throughput. The first phase is scored by a specially developed metric,
which puts high priority on efficiency of finding real hits belonging
to a particle and low fake rates. At least 50\% of the hits have to
originate from the same simulated truth particle, with hits on the
innermost layers, key for good vertex resolution, and on the outermost
layers, key for long lever arms and thus for good momentum resolution,
getting highest weights in the overall score. A random solution will
get a score of zero and a perfect reconstruction of all events in the
test dataset, consisting of 125 simulated events, will get a score of
one.

The challenge attracted more than 650 participants. In the accuracy
phase the participants provide their reconstruction of the test
dataset to Kaggle where it is scored. In the throughput phase the
participants provide their algorithms and software, and it is run in a
consistent environment (Docker containers on two i686 processor cores
and 4GB of memory) to measure both accuracy and runtime, which will be
very important to handle the enormous datasets expected from the
HL-LHC.

Winners~\cite{TrackMLWin} of the accuracy phase are teams (with
scores): top-quarks(0.92219), outrunner(0.90400) and Sergey
Gorbunov(0.89416). While training can consume lots of computer
resources, where machine learning really shines is the speed of
reconstruction once the algorithms are trained. Winners of the
throughput phase are teams sgorbuno (Sergei Gorbunov), fastrack
(Dmitry Emelyanov) and cloudkitchen (Marcel Kunze), who were able to
combine high accuracy scores with speeds well below ten seconds per
event, and even below one second for the first two.

The TrackML challenge shows that ML techniques like representation
learning, combinatorial optimization, clustering and even time series
prediction can be applied to tracking. The best solutions offer a
synergy between model-based and data-based approaches, combining the
best of both worlds: physical track models and machine learning, with
sensible trade-offs between complexity and performance.

\subsection{Particle Identification}

Particle and jet identification are examples where machine based
classification methods are rapidly replacing the traditional HEP
approaches.


The LHCb experiment at the LHC specializes in the physics of beauty
quarks. Identifying the types of long lived charged particles in the
tracker, ring-imaging Cherenkov detectors, electromagnetic and hadron
calorimeters and the muon chambers is key. Global particle
identification (PID) based on machine learning techniques is
developed~\cite{LHCbPID}. The charged particle classes are: electron,
muon, pion, kaon, proton and ghost track (fakes created by the
tracking algorithm).

The baseline PID approach, ProbNN, is based on six binary
(one-vs-rest) one-layer shallow artificial NN, implemented in the TMVA
library. Each network separates one particle type from the rest. The
DeepNN with three hidden layers of 300, 300 and 400 neurons is based
on Keras, and works in multiclassification mode to separate the six
particle types in one go. CatBoost consists of six ``gradient boosting
over oblivious decision trees classifiers'', working in one-vs-rest
mode. Sixty observables from the LHCb detectors are available for PID;
DeepNN and CatBoost use all of them, while ProbNN uses different
subsets for each PID hypotheses, based on physics reasons. The
classifiers are trained on one million simulated events for each
charged particle type.

The performance is verified on real data using kinematically
identified decays to known particles like $J/\psi \rightarrow
e^+e^-(\mu^+\mu^-)$, $\Lambda \rightarrow p\pi^-$, $D^0 \rightarrow
K^-\pi^+$. The separation quality of the different classifiers is
compared for six signal-background pairs: e-vs-$\pi$, e-vs-K,
$\mu$-vs-$\pi$, K-vs-$\pi$, p-vs-$\pi$ and p-vs-K. Different
classifiers score best for different pairs, with CatBoost and DeepNN,
by using all observables, outperforming ProbNN on most counts. The
proton-kaon separation is the most difficult, as both leave similar
traces in all detector systems. Here using all the available
information provides a clear advantage.


The Belle II experiment is operating at $\Upsilon(4S)$ center-of-mass
energy of 10.58 GeV at the SuperKEKB energy-asymmetric
electron-positron B factory with record design luminosity of
8$\cdot$10$^{35}$cm$^{-2}$s$^{-1}$, a factor of forty increase. This
will expand the intensity frontier, with the size of the Belle II
dataset expected to be fifty times bigger than the one collected by
Belle. For the study of CP violation and flavor mixing in neutral B
meson decays, the copious decays
$\Upsilon(4S)\rightarrow B^0\bar{B}^0$
are used. One of the $B$ mesons is fully reconstructed (signal side,
including all products of this decay), and the flavor of the second
(containing a b quark or antiquark) has to be determined (tag side,
the rest of the particles). This is called flavor tagging.

To ensure the success of the physics program, improved flavor taggers
using machine learning are developed~\cite{BelleIIflavor} to cope with
the ultra high luminosity and increased beam backgrounds. A
category-based tagger uses fast BDTs. A $B^0$($\bar{B}^0$) meson
contains a positively charged $\bar{b}$ (negatively charged $b$)
quark, which can decay e.g. to a positive (negative) lepton. Using
multivariate analysis, thirteen specific categories are identified,
where the flavor signatures of the measured decay particles are
correlated with the $B$ meson flavor. Each category contains one or
two particles: e, $\mu$, lepton (e or $\mu$), K, $\pi$, $\Lambda
\rightarrow p\pi^-$, called targets. PID variables from the various
subdetectors and kinematic variables (simple like momenta and impact
parameters, and global like the recoil mass) are used to identify the
targets among all the target side particles. In a first step
individual tag-side tracks are found, using 108 unique inputs. Each
particle candidate receives 13 weights in [0,1] for the probability of
being the target for a category. The candidate with the highest weight
for a category is selected as the target. The second step combines the
outputs from the thirteen categories, again using multivariate
methods. This improves the performance, as the $B^0_{tag}$ decay can
produce more than one flavor-specific signature, so more than one
category will contribute.

The performance of the category-based flavor tagger is evaluated on
simulated Belle II, and simulated and real Belle events. For the
simulated events $B^0_{sig}$ decays to $J/\psi K^0_s \rightarrow
\mu^+\mu^-\pi^+\pi^-$, while $B^0_{tag}$ has all possible decays. The
sizes of the testing and training samples are 1.3 and 2.6 (1 and 2)
million events for Belle II (Belle). Interestingly, the training
sample has to be generated {\it without} CP violation to avoid the
algorithm ``learning'' CP asymmetries on the tag side. The effective
tagger efficiency on simulated events is $\sim$37\%, a 10\% relative
improvement over the Belle result. Larger training data samples give
no further improvement. As an alternative, a deep-learning flavor
tagger, based on a multi-layer perceptron (MLP) with eight hidden
layers and 140 input variables is under development. It tries to learn
the correlations between the tag-side tracks and the $B^0_{tag}$
flavor using the full information without any preselection of decay
products. The first results are encouraging: while there is no
improvement for Belle, the Belle II results indicate progress. The
complexity of the MLP tagger requires huge training samples: the best
results so far use 55 million events for training, and the tendency is
to still improve with larger datasets. This computation takes about 48
hours with acceleration on a graphical GTX970 GPU, while the same
training consumes about five hours on a single CPU for the
category-based flavor tagger.


In many HEP measurements, identification of jet flavors is a key
component. Traditionally this is done exploiting the characteristic
features of heavy flavor charm or beauty hadrons, decaying at some
distance from the primary interaction point. This produces displaced
tracks and secondary vertices (SV), and often leptons from the sizable
leptonic and semi-leptonic branching ratios. Additional difficulties
arise from the embedding of the decay products within jets resulting
from the parton shower. At collider energies these jets can often be
highly boosted and collimated.

In the CMS collaboration the jet flavor classifier
DeepCSV~\cite{Sirunyan:2017ezt} (Combined Secondary Vertex) was
developed. It uses a dense NN of five layers with 100 nodes each with
{\it ReLU} activation, and an output layer with {\it softmax}
activation to separate four classes: b, bb (two merging B hadrons in
the jet), c and light (both quarks and gluons). The model is
implemented in Keras with a TensorFlow~\cite{TensorFlow}
backend. Sixty-eight input features enter the NN: 8 for each of the
six most displaced tracks, 8 for the most displaced SV, and 12 global
variables. Missing features are represented as zeros. Pile-up tracks,
fakes and nuclear interaction vertices are rejected in
advance. Notably, cMVAv2, the best previous tagger using additional
lepton information, was outperformed by DeepCSV.

\begin{figure}[htb]
\centerline{\includegraphics[width=\textwidth]{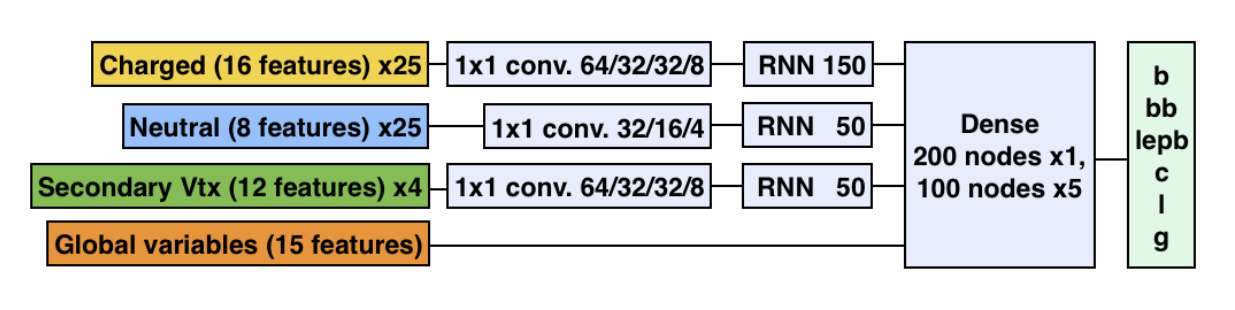}}
\caption{DeepJet network architecture.
  Image from reference~\cite{CMSJetFlavor} under a CC BY 4.0
  license (https://creativecommons.org/licenses/by/4.0/).}
\label{DeepJet}
\end{figure}

The success of deep learning in the jet arena sparked interest for
more complex models in CMS~\cite{CMSJetFlavor}, based on CNN. These
networks have been used e.g. for classification of highly boosted
jets, where the internal jet energy distribution is a major focus. The
DeepJet algorithm for flavor identification applies CNN not on images,
but on single particles. No preselection is needed. The input
variables are 16 for up to 25 displacement sorted tracks, 8 for up to
25 neutral candidates, 12 for up to 4 SV, and 15 global for a total of
up to 663 inputs. Passing through a set of convolutional layers, these
produces 8, 4, and 8 features for each input track, neutral candidate
or SV. The network automatically ``engineers'' and selects the
relevant features. This way the large number of input variables can be
handled efficiently by a ``divide-and-conquer'' strategy. The network
architecture is shown in Fig.~\ref{DeepJet}.


{\it In recurrent neural networks (RNN) the connections between the
nodes form a directed graph along a temporal sequence. The graphs
can be cyclic or acyclic. Sequences of inputs can be processed by
the same units (with same weights), giving the RNN a
``memory''. Besides helping with speech recognition, this
architecture can process inputs of variable sizes, e.g. a changing
number of tracks and jets per event. Long short-term memory (LSTM)
is a special case of RNN with feedback connections, giving it a
gated memory (state) for retaining information over longer and
better controlled time intervals.}

Three independent RNN continue the processing, producing compact
summaries of dimensionality 150, 50 and 50 for the candidate types
track, neutral or SV. These outputs are combined with the global
variables to enter a dense NN of 1 layer with 200 nodes and 5 layers
of 100 nodes each. A final output layer separates six jet classes: one
B hadron, two B hadrons, leptonic B hadron decay (three b type jets),
charm, light quark (uds) and gluon. The last layer has a {\it softmax}
activation, all the other {\it ReLU} activation. DeepJet shows sizable
improvements compared to DeepCSV, for example in $t\bar{t}$ events at
high jet \mbox{p$_T\ >\ $~90 GeV} for b jet efficiency of 90\% the
number of false positives is suppressed from 10 to 3\%. At the same
time, the light quark versus gluon discrimination is on par with
dedicated RNN binary classifiers, and slightly better than CNN using
jet images.

\subsection{Physics Measurements}

From a ML point of view the measurement of a continuous quantity is a
regression problem. The use of ML for optimizing high precision
measurements at the LHC is explored in~\cite{Bourilkov2}. The
measurement of the forward-backward asymmetry $A_{FB}$ of lepton pairs
en route to the precise determination of the electroweak mixing angle
using linear (LR) or deep NN regressors (DNNR) is developed as a test
case. The question addressed is whether neural network regressors
based on ML can perform on par (or even better) with the traditional
HEP approach, based on our knowledge of how things have been done for
decades? The beaten path of measuring $A_{FB}$ is to obtain the
cos$\theta$ distribution of the electron or negative muon in the
Collins-Soper frame~\cite{Collins:1977iv}. To this end the kinematic
variables undergo a Lorentz transformation to this frame, and then
forward and backward events are identified or a fit is performed to
extract the asymmetry, and ultimately the electroweak mixing angle.

The analysis is performed in narrow regions at different invariant
masses for the final-state lepton (dielectron or dimuon) pairs around
70, 91, 200 and 500 GeV, chosen to follow the change of the asymmetry
$A_{FB}$ with mass from negative values below the Z peak to positive
values above the peak. The event sample, simulated with the
{\tt PYTHIA}~\cite{pitia1,pitia2,pitia3} generator, consists of 3991
events after acceptance cuts for a generic LHC detector in the region
up to $|\eta|<2.4$: 75\% for training, 10\% for validation and 15\%
for testing. The input variables for the ML regressors are: dilepton
invariant mass $m$, transverse momentum $p_T$ and rapidity $y$ of the
dilepton system, pseudorapidity $\eta_{1 or 2}$ and transverse
momentum $p_T^{1 or 2}$ for each decay lepton. As the goal is to
extract the asymmetry from the decay kinematics and the angles
``hidden'' inside it, the invariant mass, which relates directly to
the forward-backward asymmetry $A_{FB}$ in narrow mass regions, is not
used. All input variables are transformed to span the range
\mbox{[-1, 1]} in order to help the minimization procedure. The target
variable is the observed forward-backward asymmetry $A_{FB}$.

Experimentation shows that the most sensitive input variables are the
rapidities. To account for the symmetric nature of the LHC, where the
observed $A_{FB}$ is stronger when the dilepton system has higher
boost away from the center, the rapidities are converted to derived
features (also called synthetic features in ML), such as the absolute
value of the rapidity $|y|$ and of the pseudorapidities $|\eta_1|$,
$|\eta_2|$. The training of the models is performed over two thousand
epochs. The performance is quantified by the root mean squared error
(RMSE) between predictions and targets for all events. A DNNR has
better chances to learn the non-linearities in the dataset at the
price of introducing higher complexity with more parameters to be
determined at the learning stage. A NN architecture well tuned to the
problem was selected: two hidden layers with ten nodes each,
{\it ReLU} activation and Adam optimizer.

For each mass region the ``measured'' $A_{FB}$ is extracted from the
ensemble of all predictions from the events in this region. To
minimize the impact of a few outliers (predictions far from the
simulated ``truth''), the median for each region is taken. The
RMSE$_{asym}$ between the so determined asymmetries and the four true
asymmetries provides a performance measure for each model training.
The traditional high energy physics way of doing the analysis by
counting forward and backward events gives RMSE$_{asym}$=0.072, the LR
0.099 and the best DNNR 0.082. After normalizing the input features
the LR provides a useful starting point. The best performance with a
DNNR is only 14\% worse relative to the traditional way, which is an
impressive result given the small size of the learning dataset for
this prototype. It is reasonable to expect that by increasing
considerably the size of the simulated samples and by tuning further
the architecture and hyperparameters of DNNR for optimal performance,
the ML results can improve and be at least on par with the traditional
approach.

\subsection{Simulations}


{\it Generative Adversarial Networks~\cite{GAN} (GAN) are a promising
recent invention. A discriminator (often a CNN) is trained to
recognize items from a training set. Then a generator (often a
deconvolutional NN) generates samples with the goal of making them
indistinguishable in statistical sense from the training set. The
two NN play an ``adversarial game'': the generator tries to fool the
discriminator by making it falsely classifying more and more
synthetic items as true. During training, the generator tries to
increase its success at fooling, while the discriminator tries to
get better at rejecting the synthetic items. At the end of the game
the true and synthetic distributions should be hard to tell apart.}

GANs have been used for the production of 2D representations of energy
depositions from particles interacting with a calorimeter - jet
images~\cite{deOliveira:2017pjk}, and for fast electromagnetic shower
simulation\cite{Paganini:2017hrr} directly producing component
read-outs in a three-dimensional multi-layer sampling LAr
calorimeter. Here attempts to apply similar techniques to tracking
detectors are reviewed.

The ALICE experiment specializes in the study of heavy-ion collisions
at the LHC. These collisions produce the most complex and highest
density events.  The prospect to use GANs for simulating clusters
(space-time points) produced by particles in the Time Projection
Chamber (TPC) of size 5x5x5 meters are explored in~\cite{AliceGAN}. A
dataset of 3D trajectories for proton-proton collisions at
center-of-mass energy 7 TeV, corresponding to the 2010 data taking, is
generated. From 300 events corresponding to one million data samples,
54872 are retained: 40000 for training and the rest for testing. Each
data sample contains the mass, charge and initial 3D momenta for a
particle, together with all the clusters in the TPC. Two types of GANs
are used: conditional Deep Convolutional GAN (condDCGAN), and
conditional LSTM GAN (condLSTMGAN). The former employs multi-layer
network with 2D convolutional/deconvolutional layers, the latter
multi-layer recurrent network with LSTM units to process recursive
data. The conditional approach adds the initial information about the
simulated particles to the network. In addition the loss function,
indicated by a plus sign in the model name, is modified to take both
the initial parameters and the deviations of the clusters from the
true trajectory into account.

The TPC resolution is between 0.8--1.2 mm. The performance of the GANs
is measured by the RMSE of the generated clusters from the trajectory
approximated as an arc (part of a helix). For the traditional HEP full
detector simulation RMSE is 1.2 mm. For a condDCGAN+ model the speedup
is 25 times with \mbox{RMSE=137 mm}, and for a condLSTMGAN+ it reaches
100 times for \mbox{RMSE=222 mm}. While the speedups are promising,
clearly a lot of work is needed to improve this prototype before
reaching a more optimal point on the speedup--accuracy curve.

\subsection{Anomaly Detection}

Quality monitoring is a key step in assuring that only good data where
all the systems perform as expected enters the physics analysis chain,
while bad data is detected and marked. At the same time the discarding
of good data is not desirable. At the LHC this involves detector and
online experts, and is time consuming. In the CMS experiment this is
done by comparing many histograms filled with critical detector
quantities to references collected at normal working points. Currently
histograms are collected over runs, which can last from seconds to few
hours. This makes certification more difficult, as transient problems
over shorter periods may be difficult to track.

A new approach for anomaly detection based on
autoencoders~\cite{CMSAnomaly} is proposed in CMS. At the LHC, a
luminosity section (LS) is defined by a fixed number of proton orbits
in the accelerator, and lasts $\sim$23 seconds. Each LS has a unique
label within the run, incrementing from number one. The goal is to
provide data certification at the LS granularity using machine
learning. In CMS 401 control histograms from the various subsystems
are monitored. Each of them is used to determine five quantiles plus
mean and standard deviation, which results in 2807 input variables for
ML. As anomalies typically are only 2\% of the dataset, it is
problematic to apply classical supervised ML classifications like
feedforward NN. Also, the failures can be of many types or new ones,
so it is difficult to collect large enough samples to train for
sizable percentage of them. To handle this a semi-supervised approach
is developed, where a model learns to detect {\it only one class - the
good one}. Even unseen future detector failures can be caught this
way!

This is implemented as an autoencoder (AE), which provides a map from
inputs to their representations through an artificial NN. First an
encoder produces a representation from the inputs. Then a decoder uses
the representation (which typically has lower dimensionality)
attempting to reconstruct the original input. An analogy is lossy
compression. Early AE applications focused on training to improve
compressed representations, but AE perform well to spot
anomalies. When the inputs deviate from the ``good'' type, the
representation suffers and the decoder output deviates from the
original, signaling problems.

\begin{figure}[htb]
\centerline{\includegraphics[width=\textwidth]{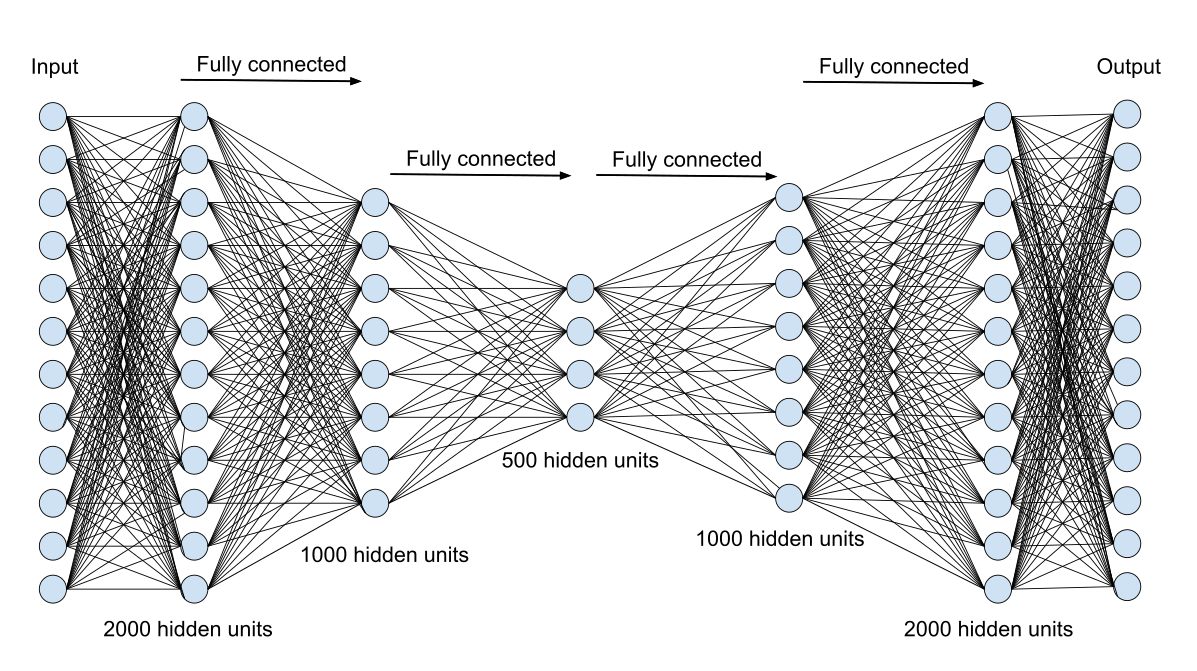}}
\caption{Autoencoder network architecture.
  Image from reference~\cite{CMSAnomaly} under a CC BY 4.0
  license (https://creativecommons.org/licenses/by/4.0/).}
\label{AutoEncoder}
\end{figure}

For this prototype 163684 samples from the 2016 data taking period are
used. They are split in 60\% for testing, 20\% for validation, and
20\% (used only once at the end) for testing. Experts removed ``by
hand'' all anomalies from the training and validation samples. The
network architecture has five hidden layers with 2000, 1000, 500, 1000
and 2000 neurons. Layers 1--3 and 3--5 correspond to the encoding and
decoding phase. The network architecture is shown in
Fig.~\ref{AutoEncoder}.

The anomalies are easily spotted by much larger MSE of
their outputs from the inputs. As an added bonus the automatic
detection also provides interpretability, as it points to the
subsystems with largest contributions to the MSE.  Future work will
focus on optimizing the model (hyperparameter tuning) and applying it
for all Run2 data and for future LHC runs.

\subsection{Ultrafast Processing for HL-LHC}

Run3 of the LHC, scheduled to begin in 2021, will bring increased
luminosities. The bunch crossing rate will be 40 MHz.

The LHCb experiment will upgrade the whole tracking system at the
heart of their physics program. An upgraded Vertex Locator (VELO) will
replace silicon strip detectors with higher resolution pixels. The
pixel size is 55 $\mu$m, starting only 5 mm from the interaction
point. This is expected to produce an excellent impact parameter
resolution of 20 $\mu$m. LHCb is embarking on a radically new
data-taking approach: a {\it trigger-less} readout, based on a
software event filter and capable to handle the 40 MHz rate. The
background suppression is dominated by the particle tracks and
displaced vertices reconstructed in the VELO.

To address this challenge, ML approaches are
developed~\cite{LHCbVELO}. The raw data for each event stores
information about all pixels that fired. Track reconstruction finds
pixels belonging to unique particle trajectories. The baseline,
traditional in HEP, algorithm starts from pixels, forms clusters of
neighbors, then track seeds, cluster doublets, track segments,
propagates them through the detector, adds more space points, etc. The
forming of doublets suffers from combinatorial explosion in the high
density LHC environments. Even starting from outside-in (from lower
density) and looking for tracks pointing to the interaction region,
the compute times are challenging. Overall, the baseline algorithm
performs well in quality and speed. But finding the optimal values for
the many tunable parameters, like search windows, is far from trivial,
and porting to GPU accelerator architectures is difficult.

A ML approach is explored to reduce the track seed combinatorics:
TrackNN, a fully connected NN (FCNN) with five layers. The non-linear
activation functions are {\it ReLU} and {\it sigmoid} for the last
layer. Acting as a classifier, for each space point on a detector
layer the FCNN finds among N space point the one with the highest
probability to form a track segment on an adjacent detector layer,
working outside-in. TrackNN reconstructs starting from the outer
layers, only retaining track extensions when the probabilities pass
quality thresholds, working towards the interaction region.

Simulated minimum bias events are used for training and testing. The
performance is measured by the reconstruction efficiency and the rate
of ``ghosts'' (spurious tracks) and ``clones'' (tracks reconstructed
more than once). For the outermost 20 planes out of 50 total, TrackNN
is almost on par with the baseline (efficiency 98.0 versus 98.9 \%),
with similar low ghost and clone rates. Moving in to using 30 and all
50 planes, the efficiency degrades to 93.2 and 71.1 \%, and the ghost
rate more than doubles. It gets harder and harder to find the real
track segments closer to the origin. So far this purely ML-based
prototype uses no {\it a-priory} knowledge about the physics of
charged particle tracks or about the detector geometry. Incorporating
this and with sustained efforts it may become possible to improve the
reconstruction quality and to benefit from substantial speed-ups
available on massively parallel architectures and on GPUs.

\subsection{Effectively Increasing the Luminosity}

An interesting aspect of the development of more powerful ML methods,
outperforming the ``traditional'' ways of doing analysis, is the
effective increase of the amount of data available to the
experiments. For example, in the CMS search for Higgs decays to
dimuons~\cite{Bourilkov1}, the net result is a 23\% increase in
sensitivity equivalent to 50\% more data. Similar result is reported
by ATLAS~\cite{ATLAS-Hmm}, where about half of the 50\% increase in
sensitivity comes from refined analysis techniques based on ML, and
the other half from increased luminosity. Perhaps on the optimistic
side, even bigger gains for specific searches are
expected~\cite{Brehmer:2018kdj} in phenomenological studies, described
in the next section.

\section{Machine Learning in Theoretical/Phenomenological
  High Energy Physics}

Building upon the sustained successes of the SM in describing the
measured phenomena in HEP, new hybrid approaches are developed pairing
the strength of cutting-edge machine learning techniques with our
knowledge of the underlying physics processes.

\subsection{Constraining Effective Field Theories}

New data analysis techniques, aimed at improving the precision of the
LHC legacy constraints, are developed in~\cite{Brehmer:2018kdj}.
Traditionally in HEP, searches for signatures of new phenomena or
limits on their parameters are produced by selecting the kinematic
variables considered to be most relevant. This can effectively explore
parts of the phase space, but leave other parts weakly explored or
constrained. By using the fully differential cross sections at the
parton level, approaches like the matrix element method or optimal
observables can improve the sensitivity in the complex cases of
multiple parameters. The weak side of these methods is how to handle
the next steps to reach the experimental data: parton showers and
detector response. Both of these steps are often simulated by
complicated Monte Carlo programs with notoriously slow convergence of
the underlying integrals. While simulations can be very accurate, they
produce no roadmap how to extract the physics from data, especially
for high dimensional problems with many observables and
parameters. Building upon our knowledge of the underlying particle
physics processes and the ability of ML techniques to recognize
patterns in the simulated data, it can be effectively summarized for
the next steps in the data analysis. In this way NN can be trained to
extract additional information and estimate more precisely the
likelihood of the theory parameters from the MC simulations.

The likelihood $\mathbf{p}(x|\theta)$ of theory parameters $\theta$
for data $x$ can be factorized in HEP as follows:
\begin{equation}
  \mathbf{p}(x|\theta) = \int dz_{detector} \int dz_{shower} \int dz \mathbf{p}(x|z_{detector}) \mathbf{p}(z_{detector}|z_{shower}) \mathbf{p}(z_{shower}|z) \mathbf{p}(z|\theta)
\end{equation}
where
$\mathbf{p}(z|\theta)\ =\ \frac{d\sigma(\theta)/dz}{\sigma(\theta)}$
is the probability density of the parton-level momenta $z$ on the
theory parameters $\theta$. The other terms in the integral correspond
to the path from partons through parton showers, detector and
reconstruction effects to the experimental data $x$ used in the
analysis. The steps on this path have the Markov property: each one
only depends on the previous one. A single event can contain millions
of variables. Calculating these integrals, and then the likelihood
function and the likelihood ratios, the preferred test statistic for
limit setting at the LHC, is an intractable problem. On the other
hand, at the parton level $\mathbf{p}(z|\theta)$ can be calculated
from the theory matrix elements and the proton parton distribution
functions for arbitrary $z$ or $\theta$ values. In this way more
information can be extracted from the simulation than just generated
samples of observables {$x$}, namely the joint likelihood ratio $r$
and the joint score $t(x,z|\theta_0)$ (which describes the relative
gradient of the likelihood to $\theta$):
\begin{equation}
  r(x,z|\theta_0,\theta_1)\ =\ \frac{\mathbf{p}(z|\theta_0)}{\mathbf{p}(z|\theta_1)}
\end{equation}
The joint quantities $r$ and $t$ depend on the parton level momenta
$z$, which for sure are not available in the measured data. Here ML
helps by using suitable loss functions based on data available from
the simulation to train a deep NN with stochastic gradient descent to
approximate functionals that can produce the important likelihood
ratio: $r(x|\theta_0,\theta_1)$ depending only on the data and theory
parameters. For technical details we refer interested readers
to~\cite{Brehmer:2018kdj} and references therein.

As a case study the weak-boson-fusion Higgs production with decays to
four leptons is taken. The {\tt RASCAL} technique uses the joint
likelihood ratio and the joint score to train an estimator for the
likelihood ratio. In essence this is a ML version of the matrix
element method, replacing very computationally intensive numerical
integrations with a regression training phase. Once the training is
complete, it takes microseconds to compute the likelihood ratio per
event and parameter point. As a bonus, the parton shower, detector and
reconstruction effects are learned from full simulations instead of
retorting to simplified, and sometimes crude, smearing functions. At
the cost of a more complex data analysis architecture, the precision
of the measurements is improved by tapping the full simulation
information. For a typical operating point from the case study, aimed
at putting limits on dimension-six operators in effective field
theories, a relative gain of 16\% on the new physics scale is
observed, corresponding to 90\% more collected data.

\subsection{Model-Independent Searches for New Physics}

So far, searches for beyond the SM (BSM), new physics (NP), phenomena
at the LHC have been negative, despite herculean efforts by the
experiments. The majority of these searches are inspired and guided by
particular BSM models, like supersymmetry or dark matter (DM). An
alternative approach, which could provide a path to NP, potentially
even lurking so far {\it unseen} in the already collected data, are
model-independent searches. They could unravel unpredicted phenomena,
for which no models are available.

One proof-of-concept~\cite{DeSimone:2018efk} strategy along these
lines is developed based on unsupervised learning, where the data are
not labeled. The goal is to compare two D (usually high) dimensional
samples: the SM simulated events (background to BSM searches), and the
real data, and to check if the two are drawn from the same probability
density distribution. If the density distributions are $p_{SM}$ and
$p_{data}$, the null hypothesis is $H_0:p_{SM}\ =\ p_{data}$, and the
alternative is $H_1:p_{SM} \neq p_{data}$. In statistical terms, this
is a two-sample test, and there are many methods to handle it. Here, a
model-independent (no assumptions about the densities), non-parametric
(compare the densities as a whole, not just e.g. means and standard
deviations) and un-binned (use the full multi-dimensional information)
two-sample test is proposed. As the densities $p_{SM}$ and $p_{data}$
are unknown, they are replaced by the estimated densities
$\hat{p}_{SM}$ and $\hat{p}_{data}$. A test statistic
(TS), based on the Kullback-Leibler KL divergence measure~\cite{KL},
is built for the ratio of the two densities, with values close to zero
if $H_0$ is true, and far from zero otherwise. The ratio is estimated
using a nearest-neighbors approach. A fixed number of neighbors K is
used, and the densities are estimated by the numbers of points within
local spheres in D dimensional space around each point divided by the
sphere volumes and normalized to the total number of points. Then the
distribution of the test statistic $f(TS|H_0)$ is derived by a
resampling method known as the permutation test, by randomly sampling
without replacement from the two samples under the assumption that
they originate from the same distribution, as expected under $H_0$.
Accumulating enough permutations to estimate the TS distribution
precisely enough, this allows to select the critical region for
rejecting the null hypothesis at a given significance $\alpha$,
e.g. 0.05, when the corresponding p-value is smaller than $\alpha$.

A proof-of-concept case study for dark matter searches with monojet
signatures at the LHC is performed. The DM mass is 100 GeV, the
mediator masses 1200--3000 GeV, detector effects are accounted for by
fast simulation, and the input features have D=8: $p_T$ and $\eta$ for
the two leading jets, number of jets, missing energy, hadronic energy
$H_T$, and transverse angle between the leading jet and the missing
energy. The comparison is done for K=5 and 3000 permutations. As an
added bonus, regions of discrepancy can be identified for detailed
scrutiny in a model-independent way. The results show promise. Before
applying them to real data, several refinements are needed: systematic
uncertainties and limited MC statistics will weaken the power of the
statistical tests, and the algorithm has to be optimized or made
completely unsupervised by automatically choosing the optimal
parameters like the value of K.

A different approach~\cite{DAgnolo:2018cun} for NP searches based on
supervised learning builds upon the same setup. This time, using the
same notation as for the unsupervised approach introduced earlier:
\begin{equation}
  p_{data}(x|\mathbf{w}) = p_{SM}(x) \cdot \exp{f(x;\mathbf{w})}
\end{equation}
where $x$ represents the d-dimensional input variables, $\mathcal{F} =
\{ f(x;\mathbf{w}), \forall \mathbf{w} \}$ is a set of real functions,
and the NP would traditionally depend on a number of free parameters
$\mathbf{w}$, introducing model dependence. Here $\mathcal{F}$ is
replaced by a neural network, in effect replacing histograms with NN,
based upon their well known capability~\cite{Cybenko} for smooth
approximations to wide classes of functions. The NP parameters are
replaced by the NN parameters, which are obtained from training on the
data and SM samples. The minimization of a suitable loss function
(which also maximizes the likelihood) provides the best fit values
$\hat{\mathbf{w}}$. Again a t-statistic and p-values are derived for
rejecting the same null hypothesis, as well as the log-ratio of the
data and SM probability density distributions.

The method is illustrated on simple numerical experiments for the
resonant and non-resonant searches for NP in the 1D invariant mass
distributions, and for a 2D case adding the $\cos{\theta}$ of the
decay products.

A limitation of these methods is the precision of the SM
predictions. Usually produced by MC full detector simulations, they
are computationally costly. In addition, systematic uncertainties of
the predictions reduce the sensitivity to new phenomena. Given the
excellent performance of the LHC and the experiments, by the end of
Run2 the data available in many corners of the phase space exceeds
the MC statistics, and the situation could get even more critical in
the future. Certainly approaches driven by data in relatively NP-free
regions, e.g. sidebands of distributions, will also have an important
role to play.

\subsection{Parton Distribution Functions}

The well known capability of NN for smooth approximations to wide
classes of functions is used in Parton Distribution Function (PDF)
fits to the available lower energy and LHC data by the
NNPDF~\cite{Ball:2014uwa,Ball:2017nwa} collaboration. The fit is based
on a genetic algorithm with a larger number of mutants to explore a
larger portion of the phase space, and nodal mutations well suited for
the NN utilized as unbiased interpolators of the various flavors of
PDFs. To avoid overfitting, the cross-validation runs over a
validation set which is never used in the training, but remembers the
best validation $\chi^2$. At the end, not the ``best'' fit on the
training set, but a ``look-back'' to the best validation fit is
retained as the final result. The NNPDF sets are easily accessible
through the LHAPDF~\cite{Bourilkov:2003kk,Whalley:2005nh,Bourilkov:2006cj,Buckley:2014ana}
libraries.

A set of Monte Carlo ``replicas'' is used to estimate the
uncertainties by computing the RMSE of predictions for physical
observables over the ensemble. In practice this works well in most
cases. Care is needed in corners of the phase space, like searches at
high invariant masses, where cross sections for some members of the
standard PDF set can become negative, or unphysical. For these cases,
a special PDF set with reduced number of replicas, but ensuring
positivity, is provided. The price to pay is enhanced PDF uncertainty
compared to other PDF families, where the PDF parameterizations
extrapolate to such phase space corners with smaller uncertainties. In
any case, comparing several families before claiming a discovery is
highly recommended.

\section{Challenges}

\subsection{Hyperparameter Tuning}

An area where the data scientist expertise is key is finding the
optimal settings and the best ML architecture for the task at
hand. Usually this requires many trials and errors before converging
on a successful approach. Ways to optimize this process with an eye on
multi core and multi GPU applications is explored in~\cite{hpt}.
Using an example for the CPU intensive process of training a GAN for a
3D calorimetry simulation, the authors first explore ways of
parallelization on multiple nodes and/or multiple GPUs, and then turn
to the task of hyperparameter optimization. Here the previously
defined loss functions are not necessarily used as a figure of merit
(FOM) to quantify the performance. Two approaches are studied:
Bayesian Optimization with a Gaussian Process prior and an
Evolutionary Algorithm. In the first one the FOM is modeled with
Gaussian processes, and by sampling the often big hyperparameter
space, successively better suggestions closer to the optimum are
discovered. In the second approach the hyperparameters are optimized
as the ``survival'' fitness of chromosomes. The initial results show
speed-ups of around 8 times for 20, and 20 times for 100
processes. The drop from linear speed-up will require additional work
to reduce the sizable overhead.

\subsection{Systematic Uncertainties}

The large amounts of data collected at colliders like the Large
Electron-Positron collider (LEP) or the LHC, and at the intensity
frontier, mean that the statistical errors on the collected data
samples tend to get quite small, and often the systematic effects
become important and even limiting. Experience shows that a large,
often dominating amount of time in data analysis is spent on
estimating and handling the systematic errors, after the express
production of first, exploratory, results.

With the wide-spread use of ML techniques, handling of the systematic
errors comes into play~\cite{Syst}. In this study simulated data from
the HiggsML challenge is used to optimize the selection of Higgs
decays in a typical supervised classification problem, where training
is done on labeled examples of simulated data. If $\mu$ is the ratio
of the number of measured signal events over the expected number, the
FOM here is the relative error of $\mu$. It depends on the true
positive (signal classified as signal) and false positive (background
classified as signal) numbers of events. The FOM has a statistical and
a systematic error. The latter comes from known nuisance factors
making the test data different form the training data, e.g. accuracy
of theoretical background predictions, detector efficiencies and
calibrations affecting simulated (training) and real (test) data
differently.

Three methods of systematic aware learning to minimize the FOM are
explored: data augmentation, pivot adversarial network (very difficult
to train), and knowledge integration. In this study the standard NN
gives the best FOM. This may be due to the fact that reaching the FOM
minimum requires to reject lots of data to obtain high signal
purities, while the systematic aware methods become robust on the
whole dataset. More work lies ahead to make them competitive.

\subsection{Sculpting}

A serious problem for the application of ML techniques in HEP is the
sculpting of variables under study, like mass or $p_T$ distributions,
which is an obstacle in a physics analysis. One example is the mass
sculpting observed when applying the DeepDoubleB
classifier~\cite{CMSJetFlavor}. To decorrelate the classifier output
from the jet mass, two Kullback--Leibler divergence penalty terms are
applied: between the original and the classifier output (``sculpted'')
distributions for the signal and for the background. In this case the
the mass decorrelation has minimal impact on the classifier
performance.

Another example ~\cite{LHCbPID} is the sculpting of PID information as
a function of kinematic variables like particle pseudorapidity, total
and transverse momentum, and event multiplicity. To reduce the
systematic uncertainty for physics measurements it is helpful to have
a flat efficiency for PID. A specific BDT Flat 4d model, utilizing a
modified loss function, is developed to guarantee the flatness of the
efficiencies.  Similar to CatBoost and Deep NN, the Flat 4d models
uses the sixty observables from all LHCb systems. The flatness comes
at the price of a somewhat decreased PID quality.

\subsection{Speed}

The trigger requirements of the HL-LHC necessitate in some cases the
deployment of field-programmable gate arrays (FPGA), integrated
circuits which can be custom configured after manufacturing. They
contain arrays of programmable logic blocks and memory elements, in
order to make them flexible for implementing e.g. ultra-fast trigger
algorithms, which would be traditionally implemented in standard
computer software running on slower well known hardware
configurations. The study~\cite{fpga} is one example of exploring the
challenges of ML with FPGA.

\section{Two-way Street: Machine Learning Meets Physics}

As demonstrated in the previous sections, ML and DL techniques can be
beneficial for a wide scope of tasks in particle physics. It works in
both directions: insights from physics can help to understand better
how ML and DL operate, and potentially to design better network
architectures. In addition, our knowledge of the underlying physics
laws and the datasets collected or simulated by the experiments
provide a more organized environment to study ML compared to the
floods of data generated e.g. by the social networks.

\subsection{Insights from Statistical Physics and Information Theory}

In~\cite{Tishby2015,Shwartz-Ziv:2017ybi} an intriguing approach to
understanding ML based on information theory and statistical physics
is explored. Similar to thermodynamics, complex many dimensional
systems are described with only a few parameters like temperature and
pressure. For a supervised network with an input, output and several
hidden layers, just two numbers summarize the network state for each
layer, relating it to the information from the input layer (encoding),
and to the labels in the output layer (decoding). In other words, each
layer compresses or extracts relevant features from the inputs and
connects them better and better to the desired targets.

For feedforward DNN with $sigmoid$ activation, interesting patterns
depending on the increasing number of epochs are observed. The
signal-to-noise ratio (SNR) exhibits 2 phases:
\begin{itemize}
  \item High SNR: high mean of gradients and small standard deviation:
    linear drop (improvement) in accuracy, memorization of input
    features
  \item Low SNR: standard deviations for gradients bigger than means;
    diffusion, random walk and forgetting irrelevant dimensions.
\end{itemize}

Modern DNN can have millions of parameters and gradients. Typically
the gradients are large in relatively few, important, dimensions, and
flat in many, irrelevant dimensions. An apt metaphor is the Rio Grande
Canyon cutting the plane near Taos in the state of New Mexico.  For
exact solutions the gradients are close to zero in the irrelevant
dimensions. The SGD method, described earlier, has noise and
variations when the batch sizes are small. As different layers hold
information about different important dimensions, these studies
propose that this actually helps. By adding noise and fluctuations to
irrelevant dimensions, ``diffusion'' helps us to forget them, filters
them out, thus aiding the convergence. SGD eliminates irrelevant
features and compresses the information for relevant features. In this
information theoretic view of layers, compression is at best weakly
dependent on the non-linearity, and happens for all kinds of
non-linear cases. This ``compression theorem'' depends strongly on the
properties of diffusion. Through the separation in relevant and
non-relevant features, the dimensionality of the problem is
effectively determined not by the number of weights to be fitted, but
by the number of relevant dimensions.

For smaller input datasets, the estimation of the covariance matrix is
distorted by noise, thus including to some extent the non-relevant
dimensions. Overfitting is determined by the local properties of the
minimum and the corresponding weights. When there is ``noise'', the
minimum is rotated somewhat from the relevant to the non-relevant
dimensions (continuing the metaphor: like small tributary canyons to
the main canyon). Diffusion for the relevant dimensions, supposedly
``protected'' by steep gradients (canyon walls), leads to decrease of
information about the labels, and to overfitting.

To summarize, the somewhat ``cavalier'' attitude of data scientists
towards minimization, developed through experimentation and intuition
as mentioned in the introduction, appears to have quite solid
foundations in statistical physics. If we ``forget'' about the
non-relevant dimensions, there is not a single ``best'' solution, but
a whole continuum of ``close enough'' solutions, which plateau in the
non-relevant dimensions. This ensures the often observed good
generalizability of the SGD method.

\subsection{Insights from Theoretical Physics for CNN}

The concepts of covariance from theoretical physics, explored by
Einstein while developing the special and general theory of
relativity, are applied in~\cite{Cheng:2019xrt} to improve the CNN
performance for image recognition. The coordinate independence was
applied first for inertial frames through Lorentz transformations, and
then generalized for local changes of coordinates which are space and
time dependent. Not surprisingly, such features are very desirable for
image recognition as sought for when developing CNN: we want the
results of identifying images to be insensitive to translations,
rotations, mirror reflections or even deformations of the same
underlying patterns. The information coming out of the network can be
equivalent for transformed images if the network architecture is
equivariant under the corresponding group action. This ideas are
inspired from renormalization group theory. In analogy to gauge
theory, the transformation laws look similar to gauge transforms, but
not quite, as they act on different space layers. Adding symmetries to
DL enables the successful use of CNN not only for the traditional
domain of two-dimensional images, but also on arbitrary manifolds,
e.g. the three-dimensional sphere as viewed by the cameras of a flying
drone.

\subsection{Interaction Networks}

A novel architecture, interaction networks~\cite{Battaglia:2016jem}
(IN), is developed by DeepMind Technologies UK, acquired by Google in
2014. The ambitious goal is to evolve methods for reasoning about
physics, objects, and their relationships, in ways bringing AI closer
to human intelligence. Capabilities like reasoning about object
interactions in complex systems, abstraction of properties and
dynamical predictions are developed. These capabilities are
challenging, because the often large amounts of inputs can be combined
in many possible, and not obvious, ways. The IN aims to do this by
building models decomposing the complexity into objects and relations,
using graphs, simulation and DL.

Objects are nodes in the network, and relationships are edges between
nodes, which can be uni- or bi-directional. Implementations based on
DNN with gradient-based optimization are possible. By separating the
objects and their interactions, many algorithms can be implemented,
without any need for ordering or pre-processing the
inputs~\cite{Battaglia:2016jem}. Case studies in the physics domain
like n-body systems, and rigid or non-rigid dynamics, show that IN
generalize well, even extending to novel, previously unseen systems
with variable configurations and numbers of objects and
relationships. In this sense a IN is a general tool showing ``elements
of reasoning''.

IN are finding promising applications in jet physics. A jet
identification algorithm~\cite{Moreno:2019bmu} is developed for the
identification of all hadronic decays of high-momentum heavy
particles. The goal is to distinguish between quark and gluon jets,
producing one collimated cone of particles, Higgs/W/Z decays producing
two, and $t \rightarrow Wq$ decays producing three overlapping cones
in the highly boosted regime. A follow-up study~\cite{Moreno:2019neq}
aims to identify boosted Higgs decays to pairs of b-hadrons.

\section{Outlook}

In this review, the many amazing manners in which AI and DL find
applications in HEP and transform the ways data is simulated or
analyzed are highlighted. By now the careful reader should be familial
with acronyms like BDT, NN, ANN, MLP, DNN, CNN, RNN, LSTM, AE, IN, SGD
and so on. These developments include all domains of physics. Building
on the trend, the American Physical Society launched this year a new
Topical Group on Data Science~\cite{APS-GDS} (GDS).

A word of caution is in order: models developed and applied
blindly as ``black boxes'' are good at learning correlations, not
necessarily at helping to understand the underlying causations. Here
physics is in an excellent position to do more by combining AI and ML
with detailed domain knowledge. The symbiosis between the two fields
is already producing encouraging results, and holds great promise for
the future. The wide adoption of machine and deep learning in the
physical sciences requires a cultural transformation and careful
strategy, combining long term goals with step-by-step solving of
current problems using the available data and human expertise, and
learning new skills in the process. The unprecedented challenges of
data-intensive science can be met with the breath-taking developments
of new artificial intelligence methods and tools.

\section*{Acknowledgments}

The author thanks his CMS colleagues from the high energy physics
group at the University of Florida for the stimulating working
atmosphere. This work is supported through NSF Grant No. PHY-1624356.



\begin{thebibliography}{00}  


\bibitem{iVDGL}
  iVDGL proposal, http://www.phys.ufl.edu/\~{}avery/ivdgl/itr2001/proposal\_all.pdf.

\bibitem{DL} 
  Y.~LeCun, Y.~Bengio and G.~Hinton,
  {\it Nature} {\bf 521}, 436 (2015).
  doi:10.1038/nature14539

\bibitem{Radovic:2018dip} 
  A.~Radovic {\it et al.},
  {\it Nature} {\bf 560}, no. 7716, 41 (2018).
  doi:10.1038/s41586-018-0361-2

\bibitem{Albertsson:2018maf}
  K.~Albertsson {\it et al.},
  {\it J.\ Phys.\ Conf.\ Ser.\ } {\bf 1085}, no. 2, 022008 (2018)
  doi:10.1088/1742-6596/1085/2/022008
  [arXiv:1807.02876 [physics.comp-ph]].

\bibitem{Carleo:2019ptp} 
  G.~Carleo, I.~Cirac, K.~Cranmer, L.~Daudet, M.~Schuld, N.~Tishby, L.~Vogt-Maranto and L.~Zdeborová,
  arXiv:1903.10563 [physics.comp-ph].

\bibitem{caltechmva}
  \mbox{{\it Workshop on Multivariate Analysis in HEP and Astrophysics}, February 11, 2008,}
  https://www.caltech.edu/campus-life-events/master-calendar/workshop-on-multivariate-analysis-in-hep-and-astrophysics-2

\bibitem{Cauchy}
  M.~Cauchy,
  {\it Comptes Rendus Hebd. S\'{e}ances Acad. Sci.} {\bf no. 25}, 536
  (1847).

\bibitem{Kingma:2014vow} 
  D.~P.~Kingma and J.~Ba,
  arXiv:1412.6980 [cs.LG].



\bibitem{Chatrchyan:2012xdj}
  S.~Chatrchyan {\it et al.} [CMS Collaboration],
  {\it Phys.\ Lett.\ B}, {\bf 716}, 30 (2012).
  doi:10.1016/j.physletb.2012.08.021
  [arXiv:1207.7235 [hep-ex]].

\bibitem{Aad:2012tfa}
  G.~Aad {\it et al.} [ATLAS Collaboration],
  {\it Phys.\ Lett.\ B}, {\bf 716}, 1 (2012).
  doi:10.1016/j.physletb.2012.08.020
  [arXiv:1207.7214 [hep-ex]].

\bibitem{Hocker:2007ht}
  A.~Hoecker {\it et al.}, arXiv:physics/0703039, 2007.

\bibitem{root}
  R.~ Brun and F.~Rademakers,
  {\it Proceedings AIHENP'96 Workshop, Lausanne, Sep. 1996,
  Nucl. Inst. and Meth. in Phys. Res. A}, {\bf 389}, 81-86 (1997).
  See also [root .cern.ch/](http://root.cern.ch/).

\bibitem{Sirunyan:2018hbu}
  A.~M.~Sirunyan {\it et al.} [CMS Collaboration],
  {\it Phys.\ Rev.\ Lett.}, {\bf 122} no.2,  021801 (2019).
  doi:10.1103/PhysRevLett.122.021801
  [arXiv:1807.06325 [hep-ex]].

\bibitem{Bourilkov1}
  D.~Bourilkov {\it et al.},
  {\it EPJ Web Conf.} {\bf 214}, 06002 (2019).
  DOI: 10.1051/epjconf/201921406002

\bibitem{ATLAS-Hmm}
  ATLAS Collab.,
  {\it ATLAS-CONF-2019-028}, CERN, July 23, 2019.

\bibitem{Chen:2016btl} 
  T.~Chen and C.~Guestrin,
  doi:10.1145/2939672.2939785
  arXiv:1603.02754 [cs.LG].

\bibitem{CMS-Hcc}
  CMS Collab.,
  {\it CMS PAS HIG-18-031}, CERN, July 14, 2019.

\bibitem{Goodfellow:2014upx} 
  I.~J.~Goodfellow {\it et al.},
  arXiv:1406.2661 [stat.ML].

\bibitem{CMS-Htautau}
  CMS Collab.,
  {\it CMS PAS HIG-18-032}, CERN, March 22, 2019.

\bibitem{CMS-ttH}
  CMS Collab.,
  {\it CMS PAS HIG-18-030}, CERN, June 24, 2019.

\bibitem{Keras}
  F.~Chollet {\it et al.}, ``Keras''. https://keras.io, 2015.

\bibitem{HiggsML}
  C.~Adam-Bourdarios {\it et al.},
  {\it JMLR: Workshop and Conference Proceedings}, {\bf 42}, 19-55, 2015.

\bibitem{HiggsMLKaggle}
  https://www.kaggle.com/c/higgs-boson/ .

\bibitem{CERN-ODP}
  CERN Open Data Portal, http://opendata.cern.ch/ .

\bibitem{Andrews}
  M.~Andrews {\it et al.},
  {\it EPJ Web Conf.} {\bf 214}, 06031 (2019).
  DOI: 10.1051/epjconf/201921406031

\bibitem{He}
  K.~He, X.~Zhang, S.~Ren and J.~Sun,
  {\it 2016 IEEE Conference on Computer Vision and Pattern Recognition},
  770 (2016).



\bibitem{BESIII}
  B.~Liu {\it et al.},
  {\it EPJ Web Conf.} {\bf 214}, 06033 (2019).
  DOI: 10.1051/epjconf/201921406033

\bibitem{GammaLearn}
  T.~Vuillaume {\it et al.},
  {\it EPJ Web Conf.} {\bf 214}, 06020 (2019).
  DOI: 10.1051/epjconf/201921406020

\bibitem{TrackML}
  M.~Kiehn {\it et al.},
  {\it EPJ Web Conf.} {\bf 214}, 06037 (2019).
  DOI: 10.1051/epjconf/201921406037

\bibitem{TrackMLKaggle}
  https://www.kaggle.com/c/trackml-particle-identification/overview .

\bibitem{TrackMLWin}
  https://sites.google.com/site/trackmlparticle/ .



\bibitem{LHCbPID}
  D.~Derkach, M.~Hushchyn and N.~Kazeev,
  {\it EPJ Web Conf.} {\bf 214}, 06011 (2019).
  DOI: 10.1051/epjconf/201921406011

\bibitem{BelleIIflavor}
  F.~Abudinén,
  {\it EPJ Web Conf.} {\bf 214}, 06032 (2019).
  DOI: 10.1051/epjconf/201921406032

\bibitem{Sirunyan:2017ezt} 
  A.~Sirunyan {\it et al.} [CMS Collaboration],
  {\it JINST} {\bf 13}, no. 05, P05011 (2018)
  doi:10.1088/1748-0221/13/05/P05011
  [arXiv:1712.07158 [physics.ins-det]].

\bibitem{TensorFlow}
  M.~Abadi {\it et al.},
  {\it TensorFlow: Large-scale machine learning on heterogeneous systems},
  2015. http://tensorflow.org/, Software available from tensorflow.org.

\bibitem{CMSJetFlavor}
  M.~Verzetti,
  {\it EPJ Web Conf.} {\bf 214}, 06010 (2019).
  DOI: 10.1051/epjconf/201921406010



\bibitem{Bourilkov2}
  D.~Bourilkov,
  {\it EPJ Web Conf.} {\bf 214}, 06022 (2019).
  DOI: 10.1051/epjconf/201921406022

\bibitem{Collins:1977iv}
  J.~C.~Collins and D.~E.~Soper,
  {\it Phys.\ Rev.\ D} {\bf 16}, 2219 (1977).

\bibitem{pitia1}
  T.~Sj\"{o}strand, S.~Mrenna and P.~Skands, {\it JHEP} {\bf 0605}, 026 (2006).

\bibitem{pitia2}
  T.~Sj\"{o}strand, S.~Mrenna and P.~Skands, {\it Comput. Phys. Comm.} {\bf 178}, 852 (2008).

\bibitem{pitia3}
  T.~Sj\"{o}strand {\it et al.}, {\it Comput.Phys.Commun.} {\bf 191}, 159-177 (2015).



\bibitem{GAN}
  I.~Goodfellow  {\it et al.},
  {\it Proceedings of the International Conference on Neural Information
    Processing Systems}, 2672–2680 (NIPS 2014).

\bibitem{deOliveira:2017pjk} 
  L.~de Oliveira, M.~Paganini and B.~Nachman,
  {\it Comput.\ Softw.\ Big Sci.\ } {\bf 1}, no. 1, 4 (2017)
  doi:10.1007/s41781-017-0004-6
  [arXiv:1701.05927 [stat.ML]].

\bibitem{Paganini:2017hrr} 
  M.~Paganini, L.~de Oliveira and B.~Nachman,
  {\it Phys.\ Rev.\ Lett.\ }  {\bf 120}, no. 4, 042003 (2018)
  doi:10.1103/PhysRevLett.120.042003
  [arXiv:1705.02355 [hep-ex]].

\bibitem{AliceGAN}
  K.~Deja, T.~Trzcin´ski and Ł.~Graczykowski
  {\it EPJ Web Conf.} {\bf 214}, 06003 (2019).
  DOI: 10.1051/epjconf/201921406003



\bibitem{CMSAnomaly}
  A.~Pol {\it et al.},
  %
  {\it EPJ Web Conf.} {\bf 214}, 06008 (2019).
  DOI: 10.1051/epjconf/201921406008



\bibitem{LHCbVELO}
  K.~Rinnert and M.~Cristoforetti,
  {\it EPJ Web Conf.} {\bf 214}, 06038 (2019).
  DOI: 10.1051/epjconf/201921406038



\bibitem{Brehmer:2018kdj}
  J.~Brehmer, K.~Cranmer, G.~Louppe and J.~Pavez,
  {\it Phys.\ Rev.\ Lett.\ } {\bf 121}, no. 11, 111801 (2018)
  doi:10.1103/PhysRevLett.121.111801
  [arXiv:1805.00013 [hep-ph]].


\bibitem{DeSimone:2018efk} 
  A.~De Simone and T.~Jacques,
  {\it Eur.\ Phys.\ J.\ C} {\bf 79}, no. 4, 289 (2019)
  doi:10.1140/epjc/s10052-019-6787-3
  [arXiv:1807.06038 [hep-ph]].

\bibitem{KL} 
  S.~Kullback, R.~Leibler,
  {\it Annals of Mathematical Statistics} {\bf 22 (1)}, 79–86 (1951).
  doi:10.1214/aoms/1177729694.

\bibitem{DAgnolo:2018cun} 
  R.~T.~D'Agnolo and A.~Wulzer,
  Phys.\ Rev.\ D {\bf 99}, no. 1, 015014 (2019)
  doi:10.1103/PhysRevD.99.015014
  [arXiv:1806.02350 [hep-ph]].

\bibitem{Cybenko}
  G.~Cybenko,
  {\it Mathematics of Control, Signals, and Systems} {\bf 2(4)}, 303–314 (1989).
  doi:10.1007/BF02551274


\bibitem{Ball:2014uwa} 
  R.~D.~Ball {\it et al.} [NNPDF Collaboration],
  {\it JHEP} {\bf 1504}, 040 (2015).
  doi:10.1007/JHEP04(2015)040
  [arXiv:1410.8849 [hep-ph]].

\bibitem{Ball:2017nwa} 
  R.~D.~Ball {\it et al.} [NNPDF Collaboration],
  {\it Eur.\ Phys.\ J.\ C} {\bf 77}, no. 10, 663 (2017).
  doi:10.1140/epjc/s10052-017-5199-5
  [arXiv:1706.00428 [hep-ph]].

\bibitem{Bourilkov:2003kk}
  D.~Bourilkov,
  hep-ph/0305126.

\bibitem{Whalley:2005nh}
  M.~R.~Whalley, D.~Bourilkov and R.~C.~Group,
  hep-ph/0508110.

\bibitem{Bourilkov:2006cj}
  D.~Bourilkov, R.~C.~Group and M.~R.~Whalley,
  hep-ph/0605240.

\bibitem{Buckley:2014ana}
  A.~Buckley {\it et al.},
  {\it Eur.\ Phys.\ J.\ C} {\bf 75}, 132 (2015).
  doi:10.1140/epjc/s10052-015-3318-8
  [arXiv:1412.7420 [hep-ph]].


\bibitem{hpt}
  J.-R.~Vlimant {\it et al.},
  {\it EPJ Web Conf.} {\bf 214}, 06025 (2019).
  DOI: 10.1051/epjconf/201921406025


\bibitem{Syst}
  V.~Estrade {\it et al.},
  {\it EPJ Web Conf.} {\bf 214}, 06024 (2019).
  DOI: 10.1051/epjconf/201921406024


\bibitem{fpga}
  A.~Tsaris {\it et al.},
  %
  {\it J. Phys. Conf. Ser.} {\bf 1085}, 042023 (2018).



\bibitem{Tishby2015}
  N.~Tishby and N.~Zaslavsky,
  {\it In Information Theory Workshop (ITW)}, {\bf 2015 IEEE}, 1–5, (2015).

\bibitem{Shwartz-Ziv:2017ybi} 
  R.~Shwartz-Ziv and N.~Tishby,
  arXiv:1703.00810 [cs.LG].

\bibitem{Cheng:2019xrt}
  M.~C.~N.~Cheng {\it et al.},
  arXiv:1906.02481 [cs.LG].

\bibitem{Battaglia:2016jem} 
  P.~W.~Battaglia, R.~Pascanu, M.~Lai, D.~Rezende and K.~Kavukcuoglu,
  arXiv:1612.00222 [cs.AI].

\bibitem{Moreno:2019bmu} 
  E.~A.~Moreno {\it et al.},
  arXiv:1908.05318 [hep-ex].

\bibitem{Moreno:2019neq} 
  E.~A.~Moreno {\it et al.},
  arXiv:1909.12285 [hep-ex].



\bibitem{APS-GDS}
  APS Topical Group on Data Science, https://www.aps.org/units/gds/index.cfm. 


\end{thebibliography}
\end{document}